\documentclass[10pt]{iopart}

\usepackage{mathrsfs}
\usepackage{iopams}
\usepackage{cite}
\usepackage{epsfig}
\usepackage{graphicx}
\usepackage{color}

\def\eps{\epsilon}
\def\Th{\Theta}
\def\sig{\sigma}

\def\3nab{\tilde{\nabla}}

\def\ra {\rangle}

\def\hsp5{\hspace{5mm}}
\newcommand{\sfrac}[2]{{\textstyle{#1\over#2}}}
\def\case#1/#2{\textstyle\frac{#1}{#2}}

\def\be {\begin{equation}}
\def\ee {\end{equation}}
\def\bea {\begin{eqnarray}}
\def\eea {\end{eqnarray}}

\def\case#1/#2{\textstyle\frac{#1}{#2} }
\def\rf#1{(\ref{#1})}

\def\equi {equilibrium\ }
\def\ra {\rightarrow }

\def\cqg{{\em Class. Quantum Grav.\/} }
\def\grg{{\em Gen. Rel. Grav.\/} }
\def\prd{{\em Phys. Rev.\/} {\bf D}}
\def\prl{{\em Phys. Rev. Lett.\/} }

\def\jmp{{\em J. Math. Phys.\/} }
\def\mn{{\em Mon. Not. Roy. Astr. Soc.\/} }
\def\aph{{\em Ann. Phys. (NY)\/} }
\def\plb{{\em Phys. Lett.\/} {\bf B}}

\begin{document}
\title[Anisotropic cosmologies with $R^n$-gravity]{Dynamical systems analysis of anisotropic cosmologies in $R^n$-gravity}
\author{Naureen Goheer \dag, Jannie A. Leach \dag \ and \\ Peter K.S. Dunsby \dag \ddag}
\address{\dag \ Department of Mathematics and Applied Mathematics, University of Cape Town, Rondebosch,
7701, South Africa}

\address{\ddag \  South African Astronomical Observatory, Observatory, Cape Town, South Africa}

\date{\today}

\eads{ngoheer@yahoo.com, jannie.leach@uct.ac.za and
peter.dunsby@uct.ac.za}

\begin{abstract}
In this paper we study the dynamics of {\it orthogonal spatially
homogeneous} Bianchi cosmologies in  $R^n$-gravity. We construct a
compact state space by dividing the state space into different
sectors. We perform a detailed analysis of the cosmological
behaviour in terms of the parameter $n$, determining all the
equilibrium points, their stability and corresponding cosmological
evolution. In particular, the appropriately compactified state space
allows us to investigate static and bouncing solutions. We find no
Einstein static solutions, but there do exist cosmologies with
bounce behaviours. We also investigate the isotropisation of these
models and find that all isotropic points are flat Friedmann like.
\end{abstract}
\pacs{98.80.JK, 04.50.+h, 05.45.-a}

\section{Introduction}
Over the past few years, there has been growing interest in higher
order theories of gravity (HOTG). This is in part due to the fact
that these theories contain extra curvature terms in their equations
of motion, resulting in a dynamical behaviour which can be different
to General Relativity (GR). In particular these additional terms can
mimic cosmological evolution which is usually associated with dark
energy \cite{DEfR}, dark matter \cite{Nojiri06,Capozzi07b} or a
cosmological constant \cite{Nojiri07}. The isotropisation of
anisotropic cosmologies can also be significantly altered by these
higher-order corrections. In a previous work \cite{Leach06}, the
existence of an isotropic past attractor within the class of Bianchi
type I models was found for a power law Lagrangian of the form
$R^n$. This feature was also found for Bianchi type I, II and IX
models in quadratic theories of gravity \cite{Barrow06c,Barrow07a}.
In these cases the extra curvature terms can dominate at early times
and consequently allow for isotropic initial conditions. This is not
possible in GR, where the shear term dominates at early times.

A natural extension of this analysis is to investigate the effect of
spatial curvature on the isotropisation in HOTG. In GR, it is
well-known that spatial curvature can source anisotropies for
Bianchi models \cite{Wald83,Goode85}. In this paper we extend the
analysis in \cite{Leach06} to the case of {\it orthogonal spatially
homogeneous} (OSH) Bianchi models \cite{vElst96}, in order to
investigate the effect of spatial curvature on the isotropisation of
$R^n$ models. OSH Bianchi models exhibit local rotational symmetry
(LRS), and include the LRS Bianchi types I (BI), III (BIII) and the
Kantowski-Sachs (KS) models. For a review of this class of
cosmologies see \cite{Ellis67,Stewart68,vElst96}).

In GR, a cosmological constant or scalar field is required to obtain
an Einstein static solution in a closed ($k=+1$)
Friedmann-Lema\^{i}tre-Robertson-Walker (FLRW) model
\cite{Goliath99,DSCosmo}.  The existence of G\"{o}del and Einstein
static universes has been investigated for gravitational theories
derived from functions of linear and quadratic contractions of the
Riemann curvature tensor \cite{Clifton05b}. Recently, the stability
of Einstein static models in some $f(R)$-theories of gravity was
investigated \cite{Bohmer07}. It was shown that the modified
Einstein static universe is stable under homogeneous perturbations,
unlike its GR counterpart \cite{Barrow03}. Static solutions are
interesting in their own right,  but are often an important first
step in finding cosmologies that have a ``bounce" during their
evolution \cite{Tolman}.

The existence conditions for a bounce to occur for FLRW universes in
$f(R)$-gravity have been determined recently \cite{Carloni06}.
Bouncing cosmological models have been found for FLRW models in
$R^n$-gravity \cite{Clifton05,Clifton07}. This should in principle
be possible for anisotropic models as well, since the higher order
corrections can mimic a cosmological constant, and so prevent the
model from collapsing to a singularity. In \cite{Solomons06}, it was
shown that bounce conditions for OSH Bianchi models   cannot be
satisfied in GR with a scalar field, but can be satisfied for KS
models in the Randall-Sundrum type Braneworld scenario.

As in \cite{Leach06}, we make use of the dynamical systems approach
\cite{Dynamical,DSCosmo,Wainwright04} in this analysis. This
approach has been applied to study the dynamics of a range of
extended theories of gravity
\cite{Carloni05,Leach06,Carloni07,Abdel07,Clifton05,Barrow06c,Carloni07a,Amendola07a,Amendola07b,Cognola07,Agarwal07}.
However, in these works, the dynamical variables were non-compact,
i.e. their values did not have finite bounds. This non-compactness
of the state space has certain disadvantages (see \cite{Goheer07b}
for detailed discussion of this issue). The standard
expansion--normalised variables for example only define a compact
state space for simple classes of ever expanding models such as the
open and flat FLRW models and the spatially homogeneous Bianchi type
I models in GR \cite{Wainwright04}. As soon as a wider class of
models or more complicated underlying gravitational theories are
considered, the expansion rate may pass through zero, making the
state space non--compact (see e.g.
\cite{Carloni05,Leach06,Carloni07,Abdel07}). The points at infinity
then correspond to a vanishing Hubble factor, and the non--compact
expansion--normalised state space can only contain the expanding (or
by time--reversal collapsing) models. In order to obtain the full
state space, one would have to carefully attempt to match the
expanding and collapsing copies at infinity.

While static solutions correspond to \equi points at infinity and
can be analysed by performing a Poincar\'{e} projection
\cite{Poincare,Perko}, bouncing or recollapsing behaviours on the
other hand are very difficult to study in this framework. In both
cases ambiguities at infinity can easily occur, since in general
only the expanding copy of the state space is studied. A point at
infinity may for example appear as an attractor in the expanding
non--compact analysis, even though it corresponds to a bounce when
also including the collapsing part of the state space.

In order to avoid these ambiguities, we will here construct compact
variables that include both expanding and collapsing models,
allowing us to study static solutions and bounce behaviour in
$R^n$-theories of gravity. This approach is a generalisation of
\cite{Goliath99}, which has been adapted to more complicated models
in \cite{Campos01a,Campos01b,Goheer04,Dunsby04}. We refer to the
accompanying work \cite{Goheer07b} for a detailed comparison between
the approach established here and differently constructed
non--compact state spaces applied to the class of BI or flat FLRW
models in $R^n$-gravity.

We note that we recover the isotropic past attractor found in
\cite{Leach06} in this analysis, and we only obtain flat ($k=0$)
isotropic \equi points. Bounce behaviour is found for BI, BIII and
KS cosmologies, but no Einstein static solutions could be found in
the phase space. Our analysis also reveals that we can have
cosmologies that bounce from expansion to contraction and vice
versa, depending on the value of the parameter $n$.

The outline of this paper is as follows: In section 2 we state the
field equations and the evolution equations for the OSH Bianchi
models. In section 3 we construct a compact state space and then
analyse the BIII and KS subspaces separately. Section 4 is devoted
to a discussion of the isotropisation of these cosmologies.

The following conventions will be used in this paper: the metric
signature is $(-+++)$; Latin indices run from 0 to 3; units are used
in which $c=8\pi G=1$.
\section{Preliminaries }

The general action for a $f(R)$-theory of gravity reads
\begin{equation}
{\cal A}=\int dx^4 \sqrt{- g} f(R)+\int {\cal L}_M dx^4,
\label{action:f(R)}
\end{equation}
where ${\cal L}_M$ is the Lagrangian of the matter fields. The
fourth order field equations can be obtained by varying
\rf{action:f(R)}:
\begin{equation}
T^M_{ab}=f' R_{ab}-\sfrac{1}{2}f g_{ab}+
S_{cd}\left(g^{cd}g_{ab}-g^c_{\; a} g^d_{\; b}\right),
\label{field:f(R)T}
\end{equation}
where primes denote derivatives with respect to $R$ and
$S_{ab}=\nabla_a \nabla_b f'(R)$. The field equations
(\ref{field:f(R)T}) can be rewritten in the standard form
\begin{equation}
G_{ab}=R_{ab}-\sfrac{1}{2}g_{ab}R=T^{eff}_{ab}, \label{field:f(R)}
\end{equation}
(when $f'(R)\neq 0$), where the effective stress energy momentum
tensor $T^{eff}_{ab}$ is given by
\begin{equation}
T^{eff}_{ab}=f'^{-1}\left[T^M_{ab}+\sfrac{1}{2}g_{ab}\left(f-f'R
\right)+ S_{cd}\left(g^c_{\; a} g^d_{\;
b}-g^{cd}g_{ab}\right)\right]. \label{field:T_tot)}
\end{equation}
It is easy to show that the contracted Bianchi identities $\nabla^a
G_{ab}=0$ give rise to the conservation laws for standard matter
\cite{Carloni05}. The propagation and constraint equations can be
obtained straightforwardly for these field equations (see
\cite{Rippl96,Leach06}).

We here consider the case $f(R)=R^n$ for OSH Bianchi spacetimes,
where the Raychaudhuri equation becomes
\begin{equation}
\dot{\Th} + \sfrac{1}{3}\,\Th^{2} + 2\sigma^2 -\frac{1}{2n}R -
(n-1)\frac{\dot{R}}{R}\Th+\frac{\mu}{nR^{n-1}} =0,
\label{Ray:R^n_BOrt}
\end{equation}
and the trace free Gauss-Codazzi equation is given by
\begin{equation}\label{sigdot:Rn_BOsh}
\dot{\sig}= -\left(\Th+
(n-1)\frac{\dot{R}}{R}\right)\sigma+\frac{1}{2\sqrt{3}}{}^3R\,.
\end{equation}
Here $\Th$ is the volume expansion which defines a length scale $a$
along the flow lines via the standard relation
$\Th=\sfrac{3\dot{a}}{a}$, and $\mu$ is the standard matter energy
density. The magnitude of the shear tensor is given by
$\sigma^2=\sfrac{1}{2}\sigma^{ab}\sigma_{ab}$, and the 3-Ricci scalar
by ${}^3R$ (see \cite{vElst96}).

The Friedmann equation is given by
\begin{equation}\label{3R:Rn_BOsh}
\sfrac{1}{3}\,\Th^{2}-\sigma^2+(n-1)\frac{\dot{R}}{R}\Th-\frac{(n-1)}{2n}R-\frac{\mu}{nR^{n-1}}+\sfrac{1}{2}\,{}^3R=0.
\end{equation}
Combining the Friedmann and Raychaudhuri equations yields
\begin{equation}\label{R:gen_BOsh}
R=2\dot{\Th}+\sfrac{4}{3}\Th^2+2\sigma^2+{}^3R.
\end{equation}
We will assume standard matter to behave like a perfect fluid with
barotropic index $w$, so that the conservation equation gives
\begin{equation}\label{cons:perfect}
\dot{\mu}=-(1+w)\mu\Th.
\end{equation}
In the following, we assume $n > 0$ and $n \neq 1$.

\section{Dynamics of OSH Bianchi cosmologies}
\subsection{Construction of the compact state space}

The overall goal here is to define compact dimensionless
expansion--normalised variables and a time variable $\tau$ such that
the system of propagation equations above
\rf{Ray:R^n_BOrt}-\rf{cons:perfect} can be converted into a system
of autonomous first order differential equations. We choose the
expansion normalised time derivative
\begin{equation}
\label{def:time-variable} '\equiv \frac{d}{d\tau}\equiv
\frac{1}{D}\frac{d}{dt}\,
\end{equation}
and make the following ansatz for our set of expansion normalised
variables \footnote{It is important to note that this choice of
variables excludes GR, i.e., the case of $n=1$. See
\cite{Dynamical,DSCosmo} for the dynamical systems analysis of the
corresponding cosmologies in GR.}:
\begin{eqnarray}
\label{def:var}
 \Sigma =\frac{\sqrt{3}\sigma}{D}\; , \quad &&x =
\frac{3\dot{R}\Th}{R D^2}(1-n)\; ,\quad y = \frac{3R}{2n D^2}(n-1)\;,\\
z = \frac{3\mu}{nR^{n-1}D^2}\; ,\quad  &&K=\frac{3 {}^3R}{2D^2}\;
,\quad \quad \quad \quad Q=\frac{\Th}{D}\,. \nonumber
\end{eqnarray}
Here $D$ is a normalisation of the form
\begin{equation} \label{def:D}
D=\sqrt{\Th^2-\Delta}\,,
\end{equation}
where $\Delta$ is a linear combination of the terms appearing on the
right hand side of the Friedmann equation \rf{3R:Rn_BOsh} as
discussed below. In order to maintain a monotonically increasing
time variable, $\Delta$ must be chosen such that the normalisation
$D$ is real--valued and strictly positive.

Note that we have chosen to define $x$ with an opposite sign to that
in \cite{Leach06} in order to have a simple form of the Friedmann
equation (see below), and $\sigma$ can be both positive and negative
\cite{vElst96}. We emphasise that the coordinates (\ref{def:var})
are strictly speaking only defined for $R\neq 0$, which means for $y\neq 0$. Even
though the case $R=0$ may not be of physical interest, the limiting
case is interesting in the context of the stability analysis, since
we obtain equilibrium points with $y=0$. This means that the system
may evolve towards/away from that singular state if these points are
attractors or repellers. In the analysis below we will investigate
this by taking the limit $y\rightarrow 0$ (by letting $R\rightarrow 0$)
and find that this puts a constraint on the relation between the
coordinates.

We now turn to the issue of compactifying the state space. It is
useful to re--write the Friedmann equation \rf{3R:Rn_BOsh} as
\begin{equation}
\label{fried-alpha} \Th^2 =
\hat{\Sigma}^2-\hat{K}+\hat{x}+\hat{y}+\hat{z}\equiv D^2+\Delta\,,
\end{equation}
where the quantities with a hat are just the variables defined in
(\ref{def:var}) without the normalisation $D$. If all the
contributions ($\hat{\Sigma}^2,~-\hat{K},~\hat{x},~\hat{y}$ and
$\hat{z}$) to the central term in equation (\ref{fried-alpha}) are
non--negative, we can simply normalise with $\Th^2$ (i.e.
$\Delta=0$), but we have to explicitly make the assumption $\Th\neq
0$. We can then conclude that the state space is compact, since all
the non--negative terms have to add up to $1$ and are consequently
bounded between $0$ and $1$.

However, while $\Sigma^2$ is always positive,
$-\hat{K},\,\hat{x},\,\hat{y}$ and $\hat{z}$ may be positive or
negative for the class of models considered here \footnote{Note that
the sign of $K$ is preserved within the open and the closed
sectors.}. This means that the variables (\ref{def:var}) do not in
general define a compact state space.

In the following, we will study the class of LRS BIII models with
${}^3R<0$ and the class of KS models with ${}^3R>0$ separately, as
in  \cite{Goliath99}. While we may in principle normalise with
$\Th^2$ in the Bianchi III subspace, we have to absorb the curvature
term into the normalisation $D$ in the KS subspace.

For both classes of models, we can construct a compact state space
by splitting up the state space into different sectors according to
the sign of $\hat{x}\,,\hat{y}$ and $\hat{z}$. In both the open and
the closed subspaces we will have to define $2^3=8$ sectors,
corresponding to the possible signs of the three variables
$\hat{x}\,,\hat{y},\,\hat{z}$. In the following, we will refer to
the spatially open BIII sectors as sector $1_o$ to sector $8_o$,
where the subscript 'o' stands for 'open'. Similarly, the spatially
closed KS sectors will be labeled sectors $1_c$ - $8_c$, where the
'c' stands for 'closed'.

After defining the appropriate normalisations for the various
sectors, we derive the dynamical equations for the accordingly
normalised variables in each sector. For each sector we then analyse
the dynamical system in the standard way: we find the equilibrium
points and their eigenvalues, which determine their nature for each
sector. The overall state space is then obtained by matching the
different sectors along their common boundaries.

\subsection{The LRS BIII subspace}
If $ {}^3R\leq 0$, we obtain the class of spatially open LRS BIII
cosmologies. This class of models contains the flat LRS BI models as
a subclass. In this case $K$ enters the Friedmann equation with a
non--negative sign and does not have to be absorbed into the
normalisation. As can be seen from the Friedmann equation in each
sector (see Table \ref{tab:norms-B3}), $K\in[-1,\,0]$ and $\Sigma\in
[-1,\,1]$ holds in each sector.
\subsubsection{Sector $1_o$}
The first open sector denoted $1_o$ is defined to be that part of
the state space where $\hat{x}\,,\hat{y},\,\hat{z}\geq 0$. In this
case all the contributions to the right-hand side of
(\ref{fried-alpha}) are non--negative, and we can choose $\Delta=0$.
This means we can normalise with $D=|\Th|=\eps\Theta$, where $\eps$
is the sign function of $\Theta$ and $\eps=\pm 1$ for
expanding/collapsing phases of the evolution. Note that it is
crucial to include $\eps$ in the normalisation: if we were to
exclude this factor, time would decrease for the collapsing models,
and any results about the dynamical behaviour of collapsing
equilibrium points would be time-reversed.

It is important to note that we have to exclude $\Th=0$ in this
sector, so we cannot consider static or bouncing solutions here.
However, this assumption is not as strong as it first appears: we
can see from the Friedmann equation (\ref{fried-alpha}) that the
only static solution in \emph{this} sector appears for
$\hat{x}=\hat{y}=\hat{z}=\hat{\Sigma}=\hat{K}=0$, because all the
quantities enter (\ref{fried-alpha}) with a positive sign in this
sector by construction. This means that we only have to exclude the
static flat isotropic vacuum cosmologies \footnote{The same
restriction appears in GR, see \cite{Goliath99}}. Under this
restriction, the normalisation above is strictly positive and thus
defines a monotonically increasing time variable via
(\ref{def:time-variable}). \noindent Equation (\ref{fried-alpha})
now becomes
\begin{equation}\label{fried-S1}
1=\Sigma^2-K+x+y+z\,.
\end{equation}
We can directly see from (\ref{fried-S1}) that the appropriately
normalised variables (\ref{def:var}) define a compact subsector of
the total state space: \be x,\,y,\,z \in [0,1]\,,~~ K\in [-1,0] ~~
\mbox{and} ~~ \Sigma \in [-1,1]\,. \ee Here $Q=\eps$ is constant and
not a dynamical variable.

This sector is different from all the other sectors in both the open
BIII and the closed KS subspaces for the following reasons. When
gluing together the different sectors to obtain the total state
space, we will actually use two copies of $1_o$: one copy with
$\eps=1$ corresponding to expanding cosmologies and one copy with
$\eps=-1$ corresponding to collapsing cosmologies. The two copies
are in fact disconnected: The closed sector $1_c$ from the KS
subspace separates the expanding and collapsing copies of open
sector $1_o$. Again, this reflects the fact that we cannot study
static solutions in sector $1_o$. In all the other sectors we allow
$\Th=0$, and the expanding and collapsing sets are connected via the
non--invariant subset $Q=0$.

We can now derive the propagation equations for the dynamical
systems variables in this sector by using the definitions (\ref{def:time-variable})
and (\ref{def:var}) and substituting them
into the original propagation equations
\rf{Ray:R^n_BOrt}-\rf{cons:perfect}. We obtain five equations, one
for each of the dynamical variables defined in (\ref{def:var}).
These variables are constrained by the Friedmann equation
(\ref{fried-alpha}), which we use to eliminate $x$, resulting in a
4--dimensional state space. Note that we have to verify that the
constraint is propagated using all five (unconstrained) propagation
equations, which we have done for each sector. The effective
system\footnote{If we used the unconstrained 5--dimensional system,
we would not constrain the allowed ranges of $n$ and $w$ for the
different \equi  points correctly. We would also get a fifth
zero--valued eigenvalue for all \equi  points.} is given by
\begin{eqnarray}\label{dyn-sys-B3}
\hspace{-12 mm} K'=2\,\eps\,K\left[1+{\Sigma}^2-\sfrac{n}{n-1}\,y
+\eps\Sigma+K\right]\,,
\nonumber\\
\hspace{-12 mm}
\Sigma'=-\eps\,\left[\eps\Sigma\left(\sfrac{2n-1}{n-1}\,y+z-2K\right)-K\right]\,,
\\
\hspace{-12 mm}y'=\frac{\eps\,y}{n-1}\left[z+(2n-3)K-(2n-1)y+(2n-1)\Sigma^2+4n-5\right]\,, \nonumber \\
\hspace{-12 mm} z' =
-\eps\,z\left[z-\Sigma^2+\sfrac{3n-1}{n-1}\,y-3K+3w-2\right]\;.
\nonumber
\end{eqnarray}
Only in this sector does the sign of the expansion--rate appears
directly in the dynamical equations, and we can see directly that
the stability of the collapsing \equi   points is given by simple
time-reversal of the stability of the expanding points and vice
versa.

The subset $K=0$ (Bianchi I) is a two dimensional  invariant
sub--manifold, so it is justified to discuss the Bianchi I subspace
on its own. This is done in detail in \cite{Goheer07b}.  The vacuum
subset $z=0$ and the submanifold $y=0$ are also invariant subspaces.
On the other hand, the isotropic subset $\Sigma=0$  is not invariant
unless $K=0$. This agrees with GR, where it was found that the
spatial curvature can source anisotropies for Bianchi models
\cite{Wald83,Goode85}.

We can find the equilibrium points and the corresponding eigenvalues
of the dynamical system (\ref{dyn-sys-B3}), and classify the
equilibrium points according to the sign of their eigenvalues as
attractors, repellers and saddle points (see \cite{Perko}). Because
of the large number of sectors that need to be studied, we do not show the
results for each sector. Instead we combine the results from the
various sectors in Table \ref{tab:eq-points}.
\subsubsection{Sectors $2_o - 8_o$}
Sectors $2_o - 8_o$ are defined according to the possible signs of
$\hat{x},\,\hat{y},\,\hat{z}$ as summarised in Table
\ref{tab:norms-B3}. In each sector $\Delta$ is defined as the sum of
the {strictly negative} contributions to (\ref{fried-alpha}), so
that $-\Delta$ is strictly positive, making $D$ strictly positive
even for $\Th=0$. This means that $D$ is a well--defined (non-zero)
normalisation, and (\ref{def:time-variable}) defines a well-defined
\emph{monotonously increasing} time variable for each sector, even
for static or bouncing solutions.
\begin{table}[!h]
\caption{Choice of normalisation in the different LRS Bianchi III
sectors, where the subscripts  in the sector labels stand for open,
differentiating the labels for the open sectors from the ones
defined in the closed KS subspace below. We abbreviate
$\hat{x}\equiv {(1-n)\dot{R}\Th}/{R},\,\hat{y}\equiv (1-n)R/{n}$ and
$\hat{z}\equiv {\mu}/{(nR^{n-1})}$.} \vspace{5mm}
\begin{tabular}{c|ccc|l|l|l}
 \br \small{sector} & \small{$\hat{x}$}&\small{$\hat{y}$}&\small{$\hat{z}$}
  &\small{normalisation} &\small{Friedmann equation}
  & \small{range of $(x,y,z)$}\\
\br
  \small{$1_o$} & \small{$\geq 0$} & \small{$\geq 0$} & \small{$\geq 0$}
  & \small{$\Delta=0$}\footnote{We must impose $\Th\neq 0$ in this sector, see text.}
  &$1=x+y+z+\Sigma^2-K$&
  \small{$[~0,~1]\times [~0,~1]\times[~0,~1]$} \\
  \small{$2_o$} & \small{$<0$} & \small{$>0$} & \small{$>0$} & \small{$\Delta=\hat{x}$}&$1=y+z+\Sigma^2-K$
  & \small{$[-1,0]\times [~0,~1]\times[~0,~1]$} \\
 \small{$3_o$} & \small{$>0$} & \small{$<0$} & \small{$>0$} & \small{$\Delta=\hat{y}$}&$1=x+z+\Sigma^2-K$&
 \small{$[~0,~1]\times [-1,0]\times[~0,~1]$} \\
 \small{$4_o$} & \small{$>0$} & \small{$>0$} & \small{$<0$} & \small{$\Delta=\hat{z}$}& $1=x+y+\Sigma^2-K$
  &\small{$[~0,~1]\times [~0,~1]\times[-1,0]$} \\
 \small{$5_o$} & \small{$<0$} & \small{$<0$} & \small{$>0$}
 & \small{$\Delta=\hat{x}+\hat{y}$} &$1=z+\Sigma^2-K$&  \small{$[-1,0]\times [-1,0]\times[~0,~1]$} \\
 \small{$6_o$} & \small{$<0$} & \small{$>0$} & \small{$<0$}
  & \small{$\Delta=\hat{x}+\hat{z}$}&$1=y+\Sigma^2-K$ & \small{$[-1,0]\times [~0,~1]\times[-1,0]$} \\
  \small{$7_o$} & \small{$>0$} & \small{$<0$} & \small{$<0$}
  & \small{$\Delta=\hat{y}+\hat{z}$}&$1=x+\Sigma^2-K$ & \small{$[~0,~1]\times [-1,0]\times[-1,0]$} \\
  \small{$8_o$} & \small{$<0$} & \small{$<0$} & \small{$<0$}
  & \small{$\Delta=\hat{x}+\hat{y}+\hat{z}$}&$1=\Sigma^2-K$& \small{$[-1,0]\times [-1,0]\times[-1,0]$} \\\br
\end{tabular}\label{tab:norms-B3}
\end{table}
With this choice of normalisation, only positive contributions
remain in the Friedmann equation, and the appropriately normalised
variables define a compact sub--sector of the total state- space, as
can be seen from the respective versions of the Friedmann equation
in Table \ref{tab:norms-B3}. Note that the Friedmann equation looks
different in each sector, which is of course due to the different
normalisation for each sector. We also gain a second constraint
equation which arises from the definition of $Q$: \be \label{con-Q}
1=Q^2-\frac{\Delta}{D^2}\,, \ee which can be written in terms of the
variables (\ref{def:var}) in each sector.

It is straightforward to derive the dynamical equations for each
sector, and again we analyse them as outlined in the previous
subsection. We confirm in each sector that the flat LRS BI subset is
indeed an invariant submanifold.
\subsubsection{Equilibrium points of the full LRS BIII state space}
The equilibrium points of the entire BIII state space are obtained
by combining the equilibrium points in each sector. We summarise
them in Table \ref{tab:eq-points}. Note that not all the points
occur in all of the sectors, and some points only occur in a given
sector for certain ranges of $n$ or a specific equation of state
$w$. For this reason, we cannot express all the equilibrium points
in terms of the same variables. When possible we state the
coordinates in terms of the dimensionless variables defined for
sector $1_o$, i.e if the given point occurs in this sector. This is
true for all the points except the line $\mathcal{L}_2$, whose
coordinates are described in terms of the variables defined in
sector $2_o$ (see below for more details on the relation between
$\mathcal{L}_{1}$ and $\mathcal{L}_{2}$).

We emphasise that if the same point occurs in different sectors, it
will have different coordinates in each of these sectors. In
particular, $Q$ can be a function of $n$ or $w$ in sectors $2_o-8_o$
even if $Q=\eps$ is a constant in sector $1_o$. This simply reflects
the fact that we have to exclude the static solutions in sector
$1_o$ but not in the other sectors. This issue will be of importance
when looking for static solutions in section \ref{subsec:static}. In
order to ensure that equilibrium points obtained in different
sectors correspond to the same solution, we have to look at the
exact solution at these points. This is outlined in section
\ref{subsec:solutions}.

Note that each of the isolated \equi points has an expanding
($\eps=1$) and a collapsing ($\eps=-1$) version as indicated in the
labeling of the points via the subscript $\eps$ in Table
\ref{tab:eq-points}. Similarly, the lines each have an expanding and
a contracting branch (see below).  We will however drop the
subscript in the following unless we explicitly address an expanding
or contracting solution.

We find the three equilibrium points $\mathcal{A}$, $\mathcal{B}$
and $\mathcal{C}$ corresponding to spatially flat Friedmann
cosmologies. The expanding versions of these points correspond to
the equally labeled points in the BI analysis \cite{Leach06} (see
\cite{Goheer07b} for detailed comparison). These points were also
found in the Friedmann analysis \cite{Carloni05}. $\mathcal{A}$ and
$\mathcal{B}$ are vacuum Friedmann points, while $\mathcal{C}$
represents a non-vacuum Friedmann point whose scale factor evolution
resembles the well known Friedmann-GR perfect fluid solution with
$a\propto t^{\frac{2}{3(1+w)}}$.

We now address the two lines of equilibrium points denoted by
$\mathcal{L}_{1}$ and $\mathcal{L}_{2}$. Both these lines correspond
to the spatially flat anisotropic BI cosmologies. The ratio of shear
$\Sigma$ and curvature component $x$ changes as we move along both
lines.  We note that in \cite{Leach06} a single line of \equi points
denoted $\mathcal{L}^*_{1}$ was found. In section \ref{Subsec:Qual}
we will discuss in more detail how $\mathcal{L}_{1}$ and
$\mathcal{L}_{2}$ are related to $\mathcal{L}^*_{1}$ .

We emphasise that for $\mathcal{L}_{1}$ the two expanding and
contracting branches are disconnected and appear as two copies
$\mathcal{L}_{1,\eps}$ of the line labeled by $\eps$ in Table
\ref{tab:eq-points}. Each of these two branches range from purely
shear dominated ($\Sigma=1$) to isotropic ($\Sigma=0$), to purely
shear dominated with opposite orientation ($\Sigma=-1$). For
$\mathcal{L}_{2}$ on the other hand the expanding and contracting
branches are connected: each $\mathcal{L}_{2,+}$ and
$\mathcal{L}_{2,-}$ ranges from expanding ($Q_\ast>0$) and static
($Q_\ast=0$) to collapsing ($Q_\ast<0$). The two disconnected copies
$\mathcal{L}_{2,+}$ and $\mathcal{L}_{2,-}$ correspond to positive
and negative values of the shear respectively. Note that there is no
isotropic subset of $\mathcal{L}_{2}$ in analogy to the fact that
there is no static subset of $\mathcal{L}_{1}$.

A closer look shows that $\mathcal{L}_{1}$ and $\mathcal{L}_{2}$ are
actually the same object in different sectors: $\mathcal{L}_1$ has
$\hat{x}\geq 0$ hence occurs in sectors $1_o$, $3_o$, $4_o$ and
$7_o$, while $\mathcal{L}_2$ is the analog with $\hat{x}<0$
occurring in sectors $2_o$, $5_o$, $6_o$ and $8_o$. This statement
is confirmed by looking at the exact solutions corresponding to the
points on both lines; we find that both these lines have the same
parametric solution of scale factor and shear (see section below).
For this reason, we could in fact give the two lines the same label.
However, it is useful to treat them separately, since we obtain
different bifurcations in the sectors with $\hat{x}>0$ and
$\hat{x}<0$ respectively. Furthermore, the subset of the line
denoted by $\mathcal{L}_2$ allows for static solutions unlike the
subset labeled $\mathcal{L}_1$. This is due to the fact that a
negative curvature contribution $\hat{x}$ can effectively act as a
cosmological constant by counter--balancing other contributions in
the Friedmann equation. This is explored in section
\ref{subsec:static} below.

Finally, we find the \equi points $\mathcal{D}$ and $\mathcal{E}$
corresponding to spatially open models. Point $\mathcal{D}$ is
independent of $n$ and $w$, while $\mathcal{E}$ depends on the value
of $w$. The points $\mathcal{F}$ and $\mathcal{G}$ can be spatially
open, flat or closed depending on the value of $n$ and/or $w$, i.e.
they move through the different sectors of the total state space as
$n,~w$ are varied. This is reflected in Tables \ref{tab:nature-B3}
and \ref{tab:nature-KS}, where we summarise the stability properties
of the equilibrium points of the closed and open subspaces
separately, and observe that these two points occur in each subspace
for certain ranges of $n$ only.
\begin{table}[tbp] \centering
 \caption{Equilibrium points of the full OSH Bianchi state space
 in terms of the coordinates defined for sector $1_0$, except for the line $\mathcal{L}_{2,\pm}$, where we
have to use the coordinates defined for BIII sector $2_o$ (see
text). Here $\eps=\pm 1$ labels the expanding/contracting solutions.
We have abbreviated $f(w)=\sfrac{-3(3w^2-3w+1)}{(3w-2)^2}$,
$P_1(n)=2n^2-2n-1$ and $P_2(n)=2n^2-5n+5$. We will not explicitly
state the expressions for $R_1(n,w),...,R_4(n,w)$, which are
rational functions of $n$ and $w$. The constants
$Q_\ast,~\Sigma_\ast$ take real values in $[-1,\,1]$. We have
denoted the coordinates $Q$ that can become non-constant in sectors
other than the first sector with the superscript $\dag$. }

\begin{tabular}{lll}
\multicolumn{3}{c}{}\\
\br Point  & $(Q,K,\Sigma,x,y,z)$ & Description
\\ \hline
$\mathcal{A}_\eps$ & $\left(\eps^\dag
,\;0,\;0,\;\sfrac{2(2-n)}{2n-1},\;\sfrac{4n-5}{2n-1},\;0\right)$ & Friedmann flat\\

$\mathcal{B}_\eps$ & $\left(\eps^\dag,\;0,\;0,\;3w-1 ,\;0,\;
2-3w\right)$ &
Friedmann flat\\

$\mathcal{C}_\eps$ &
$\left(\eps^\dag,\;0,\;0,\;\frac{3(n-1)(1+w)}{n},\;
\frac{(n-1)[4n-3(1+w)]}{2n^2},\frac{n(13+9w)-2n^2(4+3w)-3(1+w)}{2n^2}\right)$ & Friedmann flat\\

Line $\mathcal{L}_{1,\eps}$ &
$\left(\eps,\;0,\;\Sigma_*,\;1-\Sigma_*^2,0,0\right)$ & flat LRS Bianchi I\\

Line $\mathcal{L}_{2,\pm}$ & $\left(Q_*,\;0,\;\pm
1,\;Q_*^2-1,0,0\right)$
& flat LRS Bianchi I\\

$\mathcal{D}_\eps$ &
$\left(\eps,\;-3/4,\;-\eps/2,\;0,\;0,\;0\right)$ & open LRS BIII\\

$\mathcal{E}_\eps$ &
$\left(\eps^\dag,\;f(w),\;\sfrac{\eps}{3w-2},\;3w\sfrac{3w-1}{3w-2},\;0,\;3w\,f(w)\right)$ & open LRS BIII\\

$\mathcal{F}_\eps$ &
$\left(\eps^\dag,\;\frac{3(4n^2-10n+7)P_1(n)}{{P_2(n)}^2},\eps\frac{P_1(n)}{P_2(n)}
,\;\frac{6(n^2-3n+2)}{P_2(n)},\;\frac{9(4n^4-18n^3+31n^2-24n+7)}{P_2(n)^2},\;0\right)$ & vacuum BI, BIII or KS\\

$\mathcal{G}_\eps$ & $\left(\eps^\dag,\;R_1(n,w),\;\eps
R_2(n,w),\;\frac{3(n-1)(1+w)}{n},\;
R_3(n,w),\;R_4(n,w)\right)$ & BI, BIII or KS \\

& &\\
\br
\end{tabular}\label{tab:eq-points}
\end{table}

\subsection{The Kantowski-Sachs subspace}

When $ {}^3R> 0$, we obtain the class of spatially closed KS
cosmologies. Here $\hat{K}$ is positive and needs to be absorbed
into the normalisation in all sectors. This means that in this
subspace, $-\Delta$ is strictly positive in all closed subsectors
$1_c$ - $8_c$, hence $D^2$ is strictly positive even for $\Th=0$. We
can therefore consider static and bouncing solutions in all sectors
that make up the closed subspace. The flat subspace is obtained in
the limit ${}^3R\rightarrow 0$. As explained in the previous
subsection, we have to exclude static flat isotropic vacuum
cosmologies in this limit.

The closed sectors can be defined as in the BIII case, except that
$K$ no longer appears in the Friedmann equation  (see Table
\ref{tab:norms_KS}). Similar to the BIII case, the first sector
labeled $1_c$ is defined as the subset of the state space where
$\hat{x}\,,\hat{y},\,\hat{z}\geq 0$. In this case we choose
$\Delta=-\hat{K} (<0)$, so that equation (\ref{fried-alpha}) becomes
\begin{equation}\label{fried-S1_KS}
1=\Sigma^2+x+y+z\,.
\end{equation}
The curvature can be obtained from (\ref{con-Q}), which in this
sector becomes
\begin{equation}\label{cont:QK}
1=Q^2+K.
\end{equation}
From (\ref{fried-S1_KS}) and (\ref{cont:QK}) it is clear that the
appropriately normalised variables (\ref{def:var}) define a compact
subsector of the total state space with
\begin{equation} x,\,y,\,z \in
[0,1]\,, ~~ K \in [0,1] ~~\mbox{and} ~~ Q,\,\Sigma \in [-1,1]\,.
\end{equation}
Note that the variable $K$ will not be used explicitly in any of the
closed sectors.

As in the BIII case, we derive the propagation equations for the
dynamical systems variables in this sector and reduce the
dimensionality of the state space to four by eliminating $x$ via the
Friedmann constraint (\ref{fried-S1_KS}). Again we have verified
that the constraint is preserved using all five propagation
equations. We obtain the following dynamical system:
\begin{eqnarray}\label{dyn-sys-KS}
\hspace{-15 mm} Q'= \frac{1}{3}(Q^2-1)\left[1+Q\Sigma+
\Sigma^2-\frac{ny}{n-1}\right] \, ,\nonumber  \\
\hspace{-15 mm} \Sigma'=\frac{\Sigma}{3Q}\left[Q^2\left(
\Sigma^2-1-\frac{ny}{n-1}\right)+1-\Sigma^2-y-z\right]+\frac{1}{3}(\Sigma^2-1)(Q^2-1)\,
,
\\
\hspace{-15 mm}y'=\frac{y}{3Q}\left[2\Sigma
Q(Q^2-1)-\frac{1}{n-1}(1-\Sigma^2-y-z)+2Q^2\left(2+\Sigma^2
-\frac{ny}{n-1}\right)\right]\,, \nonumber \\
\hspace{-15 mm} z' = \frac{z}{3Q}\left[2\Sigma
Q(Q^2-1)+1-\Sigma^2-y-z+Q^2\left(1-3w+2\Sigma^2
\frac{-2ny}{n-1}\right)\right]\,. \nonumber
\end{eqnarray}
We recover the following features from the BIII subspace: The flat
subset $K=0$ (here corresponding to $Q^2=1$) is invariant, as can be
seen from the $Q'$-equation together with the Friedmann equation
(\ref{fried-S1_KS}). Other invariant subspaces are the
hyper-surfaces $y=0$ and $z=0$. The isotropic subset $\Sigma=0$ is
not invariant unless $K=0$.
\begin{table}[!h]
\caption{Choice of normalisation in different KS sectors, where the
subscripts in the sector labels stand for closed. See text and
caption of Table \ref{tab:norms-B3} for details on the notation used
here.} \vspace{5mm}
\begin{tabular}{c|ccc|l|l|l}
 \br \small{sector} & \small{$\hat{x}$}&\small{$\hat{y}$}&\small{$\hat{z}$}
  &\small{normalisation} &\small{Friedmann equation}
  & \small{range of $(x,y,z)$}\\
\br
  \small{$1_c$} & \small{$\geq 0$} & \small{$\geq 0$} & \small{$\geq 0$}
  & \small{$\Delta=-\hat{K}$}&$1=x+y+z+\Sigma^2$&
  \small{$[~0,~1]\times [~0,~1]\times[~0,~1]$} \\
  \small{$2_c$} & \small{$<0$} & \small{$>0$} & \small{$>0$} & \small{$\Delta=\hat{x}-\hat{K}$}&$1=y+z+\Sigma^2$
  & \small{$[-1,0]\times [~0,~1]\times[~0,~1]$} \\
 \small{$3_c$} & \small{$>0$} & \small{$<0$} & \small{$>0$} & \small{$\Delta=\hat{y}-\hat{K}$}&$1=x+z+\Sigma^2$&
 \small{$[~0,~1]\times [-1,0]\times[~0,~1]$} \\
 \small{$4_c$} & \small{$>0$} & \small{$>0$} & \small{$<0$} & \small{$\Delta=\hat{z}-\hat{K}$}& $1=x+y+\Sigma^2$
  &\small{$[~0,~1]\times [~0,~1]\times[-1,0]$} \\
 \small{$5_c$} & \small{$<0$} & \small{$<0$} & \small{$>0$}
 & \small{$\Delta=\hat{x}+\hat{y}-\hat{K}$} &$1=z+\Sigma^2$&  \small{$[-1,0]\times [-1,0]\times[~0,~1]$} \\
 \small{$6_c$} & \small{$<0$} & \small{$>0$} & \small{$<0$}
  & \small{$\Delta=\hat{x}+\hat{z}-\hat{K}$}&$1=y+\Sigma^2$ & \small{$[-1,0]\times [~0,~1]\times[-1,0]$} \\
  \small{$7_c$} & \small{$>0$} & \small{$<0$} & \small{$<0$}
  & \small{$\Delta=\hat{y}+\hat{z}-\hat{K}$}&$1=x+\Sigma^2$ & \small{$[~0,~1]\times [-1,0]\times[-1,0]$} \\
  \small{$8_c$} & \small{$<0$} & \small{$<0$} & \small{$<0$}
  & \small{$\Delta=\hat{x}+\hat{y}+\hat{z}-\hat{K}$}&$1=\Sigma^2$& \small{$[-1,0]\times [-1,0]\times[-1,0]$} \\\br
\end{tabular}\label{tab:norms_KS}
\end{table}

The sectors $2_c$-$8_c$ are defined according to the possible signs
of $\hat{x},\,\hat{y},\,\hat{z}$ as summarised in Table
\ref{tab:norms_KS}: In each sector $\Delta$ is defined as the sum of
the {strictly negative} contributions to (\ref{fried-alpha}). The
dynamical equations analogous  to \rf{dyn-sys-KS} can be derived
straightforwardly for each sector. We then solve these equations in
each sector for their respective equilibrium points and the
corresponding eigenvalues, and classify the \equi  points according
to their dynamical properties. The results are combined with the
results from the open sectors and summarised in Tables
\ref{tab:nature-B1}--\ref{tab:nature-KS}.
\subsection{Exact solutions corresponding to the \equi points}
\label{subsec:solutions}
We now derive the solutions corresponding to the various \equi
points. Special attention has to be paid to the points with $y=0$,
since these correspond to the limit $R\rightarrow 0$, which may makes
the coordinate $x$ singular. We will study this issue in detail
below. Note that it is legitimate to take the limit $R\rightarrow 0$
in the original field equations as long as $n>1$, which results in
the constraint $T^M_{ab}\ra 0$. Consequently it is only possible to
study the limit $R\rightarrow 0$ for $\mu \ra0$ and $n>1$ when
solving for the solutions corresponding to the \equi  points with
$y=0$.

It is important to emphasise that the dynamical system by itself is
well-defined for $y=0$; only when going back to the original
equations to solve for the exact solutions corresponding to the
\equi  points with $y=0$ do we notice that there may not be an exact
solution corresponding to these coordinates.


We now proceed to find the exact solutions corresponding to the
non--static ($\Theta\,,Q\neq 0$) \equi  points. As usual, we can solve
the energy conservation equation (\ref{cons:perfect}) for the
non--vacuum solutions to obtain
\begin{equation}\label{mu}
\mu=\mu_0\,a^{-3(1+w)}\,,
\end{equation}
where $\mu_0$ is determined by the  $z$-coordinate of the given
equilibrium point. We require $\mu_0\geq 0$, which constrains the
allowed range of $n$ or $w$ for a given \equi  point (see below).

In order to determine the scale factor evolution at each \equi
point, we rewrite the Raychaudhuri equation \rf{Ray:R^n_BOrt} as
\begin{equation}\label{Ray:q}
\dot{\Th}=-\left(1+q_i\right)\frac{\Th^2}{3}\,,
\end{equation}
where we express the deceleration parameter $q_i$ at each point in terms of the
dimensionless variables \rf{def:var}:
\begin{equation}\label{DS:q_Osh}
q_i=2\frac{\Sigma_i^2}{Q_i^2}+\frac{x_i}{Q_i^2}-\frac{y_i}{(n-1){Q_i^2}}+\frac{z_i}{Q_i^2}\,.
\end{equation}
Note that this equation is invariant in different sectors: for a given \equi  point,
each coordinate divided by ${Q^2}$ is the same in all sectors. This ensures that the
corresponding solution is invariant, no matter with which coordinates we describe the \equi  point.

Similarly, we re--write the trace free Gauss Codazzi equation \rf{sigdot:Rn_BOsh} as
\begin{equation}
\label{DS:shear_Osh} \dot{\sig}=
-\frac{1}{\sqrt{3}Q_i^2}\left[\left(Q_i-\frac{x_i}{3Q_i}\right)\Sigma_i-\frac{K_i}{3}\right]\,
\Th^2,
\end{equation}
and the curvature constraint \rf{R:gen_BOsh} as
\begin{equation}\label{DS:R}
R=\frac{2}{3}\Theta^2\left[1-q_i+\frac{\Sigma_i^2}{Q_i^2}-\frac{K_i}{Q_i^2}\right]
\end{equation}
for a given \equi point with coordinates $(Q_i, K_i, \Sigma_i, x_i,
y_i, z_i)$ and deceleration parameter  $q_i$.
\subsubsection{Power-law solutions}
\label{}

We first study the non--stationary ($q\neq -1$) cosmologies, for
which \rf{Ray:q} has the solution
\begin{equation}
\label{sol-Theta} \Theta=\frac{3}{(1+q_i)\,t}\,.
\end{equation}
We have set the Big Bang time $t_0=0$. Given $\Theta$, we can solve
for all the other dynamical quantities for a given \equi  point to
obtain the scale factor evolution
\begin{equation}
\label{DS:sol_Osh} a=a_0\;|t|^{\alpha},\ \ \ {\rm where} \ \ \
\alpha= \left(1+q_i\right)^{-1}\,,
\end{equation}
the shear
\begin{equation}
\label{DS:shear_sol_Osh}
\sigma=\frac{\beta}{t}+const\,,~~~\mbox{where}~~
\beta=\frac{\sqrt{3}}{(1+q_i)^2}\left[\frac{\Sigma_i}{Q_i}\left(3
-\frac{x_i}{Q_i^2}\right)
 - \frac{K_i}{Q_i^2}\right]\,,
\end{equation}
and the curvature scalar
\begin{equation}
\label{R:sol_fix} R=\frac{\gamma}{t^{2}}\,,~~~\mbox{where}~~
\gamma=\frac{6}{(1+q_i)^2}\left[1-q_i+\frac{\Sigma_i^2}{Q_i^2}+\frac{K_i}{Q_i^2}\right]\,.
\end{equation}

Again, we point out that even though a given \equi  point formally
has different coordinates in the different sectors, the exact
solutions corresponding to the point are invariant, since the
coordinates only enter the solutions \rf{DS:sol_Osh}--\rf{R:sol_fix}
with a factor ${1}/{Q_i^2}$. The solutions for each point are
summarised in Table~\ref{tab:sols}, where constants of integration
were obtained by substituting the solutions into the original
equations.

When substituting the points with $y=0$ into the original field
equations, we find that these are only satisfied for special values
of $n$. This is reflected in Table \ref{tab:sols}. Point ${\mathcal
B}$ only has a solution for $n=5/4$ and $w=2/3$. The solutions for
points ${\mathcal D}$ and ${\mathcal E}$ only satisfy the original
equations for $n=1$, which has been excluded from the start. These
points therefore do not have any physical power--law solutions. The
points on the lines  ${\mathcal L}_{1,2}$ only have corresponding
solutions for special coordinate values, making only two points on
each line physical (see below).

Excluding these non-physical points, we find that the only
non--vacuum solutions are given by $\mathcal{C}$ and $\mathcal{G}$.
Substituting the solution \rf{mu} into the definition of $z$, we
find that the constant $\mu_0$ must satisfy
\begin{eqnarray*} \mu_0 &=& z_i y_i^{n-1}\left(\frac{\alpha}{Q_i}\right)^{2n}(3n)^n\left(\frac{2}{n-1}
\right)^{n-1}.
\end{eqnarray*}
In order for these solutions to be physical, we require that $\mu>0$
and therefore $\mu^i_0>0$. For $\mathcal{C}$ we find that this
condition is satisfied for
\begin{equation}
1<n<\frac{1}{4(4+3w)}\left(13+9w +\sqrt{9 w ^2+66 w +73}\right)\;,\
\ \ w> -1,
\end{equation}
while for $\mathcal{G}$ it is only valid for
\begin{equation}
\left\{ \begin{tabular}{ll} $1<n<N_+$\;,& \ \ \ $-1<w\leq 0$\, ,\\
$N_- <n<N_+$\;,& \ \ \ $0<w< \sfrac{1}{15}(-15+4\sqrt{15})$\,
,\end{tabular}\right.
\end{equation}
where $N_\pm=\frac{1}{4(2+w)}\left(9+5w \pm
\sqrt{1-30w-15w^2}\right)$.

For points $\mathcal{A}$ and $\mathcal{F}$ the solutions only depend
on $n$, while the solutions at $\mathcal{C}$ and $\mathcal{G}$
depend on both $n$ and $w$. We can see from these solutions that
points $\mathcal{A}$ and $\mathcal{C}$ are the isotropic analogs of
points $\mathcal{F}$ and $\mathcal{G}$ respectively.

The lines $\mathcal{L}_{1}$ and $\mathcal{L}_{2}$ have the same
solutions for shear and energy density. As noted above, they are the
same line but for different ranges of $\hat{x}$ and hence
$\alpha$. $\mathcal{L}_{1}$ contains the isotropic subset of
solutions ($\Sigma_*=0$) while $\mathcal{L}_{2}$ contains the static
subset ($\alpha=Q_*=0$).

In Table \ref{tab:q} we summarise the behaviour of the deceleration
parameter $q$. By studying the deceleration parameter, we can
determine whether the power law solutions above correspond to
accelerated ($-1<q<0$) or decelerated ($q>0$) expansion or
contraction.  The expansion (or contraction) of point $\mathcal{A}$
is decelerating for $n\in(0,1/2)$ or
$n\in(1,\sfrac{1}{2}(1+\sqrt{3}))$ and accelerating for
$n\in(\sfrac{1}{2}(1+\sqrt{3}),2)$. Point $\mathcal{B}$ and lines
$\mathcal{L}_{1,2}$ only admit decelerating behaviours. Point
$\mathcal{F}$ has a decelerated behaviour for
$n\in(0,\sfrac{1}{2}(1+\sqrt{3}))$ and an accelerated behaviour for
$n\in(\sfrac{1}{2}(1+\sqrt{3}),2)$. The \equi points $\mathcal{C}$
and $\mathcal{G}$ for $w\in [0,1]$, have decelerated behaviours when
$n\in(0,\sfrac{3}{2}(1+w))$ and accelerated behaviours when
$n\in(\sfrac{3}{2}(1+w),\infty)$.
\subsubsection{Stationary solutions}
If $q=-1$, we obtain stationary solutions ($\dot{\Theta}=0$), which
have an exponentially increasing scale factor. As reflected in Table
\ref{tab:q}, the vacuum points ${\mathcal A}$ and ${\mathcal F}$
correspond to de Sitter solutions for the bifurcation value $n=2$
for all equations of state, while the matter points ${\mathcal C}$
and ${\mathcal G}$ are de Sitter--like for all $n>0$ but $w=-1$
only, and ${\mathcal E}$ appears to be de Sitter--like for $w=1$ for
all values of $n>0$. Since ${\mathcal E}$ has $y=0$, we will have to
study this case in more detail below.

For a constant expansion rate
\begin{equation}
\Th=\Th_0\,,
\end{equation}
the scale factor has the following solution
\begin{equation}\label{DS:sol_Osh2}
a=a_0\;e^{\frac{1}{3} \Th_0 t}.
\end{equation}
The energy conservation equation becomes
\begin{equation}
\dot{\mu}=0 \Rightarrow{\mu}=\mu_0\,.
\end{equation}
The trace free Gauss Codazzi equation \rf{sigdot:Rn_BOsh} can  be
rewritten as
\begin{equation}\label{DS:shear_Osh2}
\dot{\sig}=\beta_0, \ \ \ \ {\rm where}\ \ \ \ \beta_0=
\frac{\Th_0^2}{3\sqrt{3}Q_i^2}\left[K_i-\left(3Q_i-\frac{x_i}{Q_i}\right)\Sigma_i\right],
\end{equation}
which on integration yields
\begin{equation}\label{DS:shear_sol_Osh2}
\sigma=\beta_0 t+\sigma_0\;,
\end{equation}
where $\sigma_0$ is an integration constant. The evolution of the
Ricci scalar can be obtained by substituting the solutions above
into \rf{R:gen_BOsh}, to find
\begin{equation}\label{R:sol_fix2}
R=\frac{2}{3}\left(2+\frac{K_i}{Q_i^2}\right)\Th_0^2+2 (\beta_0
t+\sigma_0)^2.
\end{equation}
As before, we substitute the solutions at each \equi point into the
definition of the coordinates, which constrains the constants of
integration for each point. In particular, $\beta_0=0$ holds for all
stationary equilibrium points, which means that we only have
constant or vanishing shear.

As in the power--law case, we see that all the \equi points except
for $\mathcal{C}$ and $\mathcal{G}$ correspond to vacuum solutions
$\mu=0$. For point $\mathcal{C}$ the energy density is given by
\begin{equation}\label{DS_Osh:mu_dS_C}
\mu=\mu^\mathcal{C}_1=4^{n-1}3^{-n}(2-n)\Th_0^{2n},
\end{equation}
and for $\mathcal{G}$ the energy density is given by
\begin{equation}\label{DS_Osh:mu_dS_G}
\mu=\mu^\mathcal{G}_1=4^{n-1}(2-n)\Th_0^{2n}.
\end{equation}
Both of these solutions only hold for $1<n\leq 2$ with $w=-1$.

Again, we substitute the generic solutions into the original field
equations for each point, and find that the original equations are
satisfied for all points with $y\neq 0$. It is however not possible
to find a stationary solution at point ${\mathcal E}$ (which has
$y=0$), even after carefully considering the limit $y\ra 0$.

\subsubsection{Static solutions}
\label{subsec:static}
The static \equi points are characterised by
$\Theta=\dot{\Theta}=0$. These points satisfy $Q=x=0$, where the
second identity comes from the fact that if $Q=0$, then we require
that $x=0$ from the definition of the variables, as discussed below.
\footnote{Note that unlike in the bouncing or recollapsing case
below, we do not consider $Q=y=0,~x\neq 0$ here, since this
corresponds to the limit $R\rightarrow 0$. While we may want to
study a bounce where the Ricci scalar approaches zero and then grows
again, we are not interested in static solutions that have vanishing
Ricci curvature at \emph{all} times.}

We will now explore which of the equilibrium points obtained above
correspond to static solutions. As indicated above, even though
$Q=\pm\eps$ holds in the first sector as stated in Table \ref{tab:eq-points},
$Q$ can be a function of $n$ and/or $w$ in the other sectors. In order to find
the static equilibrium points, we have to look at the coordinates
that each \equi  point takes in each sector, and find the values of
$n$ and/or $w$ for which $Q=0$ in the given sector.

An obvious static solution appears to be the subset $Q=0$ on line
${\mathcal L}_{2,\pm}$ for all values of $n$ and $w$. We can however
not find a solution corresponding to this limit, since $Q_\ast=0$
implies $\sigma=0$, which contradicts the value of the shear
coordinate of this \equi  point. We can study the eigenvalues
associated with the line ${\mathcal L}_2$ in the limit $Q\rightarrow
0^\pm$ and find that the static subset is an unstable saddle point
for all values of $n$ for both ${\mathcal L}_{2,+}$ and ${\mathcal
L}_{2,-}$.

The point ${\mathcal A}$ appears to admit a static solution for the
bifurcation value $n=1/2$. This bifurcation only occurs in sectors
2, 3, 6 and 7 of the open and the closed sectors. However, it is not
possible to find a solution satisfying the coordinates of the static
\equi  point that satisfies the original field equations. For this
reason, this static \equi  point is unphysical. We explore the
stability of the static solution in the limit $n\rightarrow 1/2$
from the appropriate sides: for example, point ${\mathcal A}$ only
lies in the open sector 2 for $n\in [0,\,1/2]$ or $n\in
[2,\,\infty]$, making only the limit $n\rightarrow 1/2^-$
well-defined. We find that this bifurcation represents a saddle
point in the state space since two of the eigenvalues approach
$\infty$ from the left and $-\infty$ from the right, making the
point unstable.

Even though the $Q$--coordinate of point ${\mathcal B}$ is a
function of $w$ in sectors 2, 4, 5 and 7, $Q$ cannot be zero for any
values of $w$. This means point ${\mathcal B}$ does not admit any
static solutions.

Point ${\mathcal C}$ can only be static in the limit $n\rightarrow
0$ in sector 6 for $w=0,\,1/3,\,1$ and in sector 3 and 5 for $w=-1$.
Again, we cannot find a solution for this special case, but this
case is physically not interesting either way.

The $Q$--coordinate of point ${\mathcal{E}}$ is zero in sectors 6-8
for $w=2/3$, but again there is no solution corresponding to this
limit.

Even though point ${\mathcal F}$ has $Q$ as a function of $n$ in
open sectors 2 and 6, $Q(n)$ is non--zero for the allowed ranges of
n.

Point ${\mathcal G}$ becomes static in the limit $n\rightarrow 0$ in
sectors 4, 6 and 8, which again is not physically relevant.


\begin{table}[tbp] \centering
\caption{Solutions for scale factor, shear, curvature and energy
density corresponding to the \equi points.}
\begin{tabular}{llllll}
\multicolumn{6}{c}{}\\
\br Point  & &  Scale factor ($a$) & Shear ($\sigma$)  & Ricci
Scalar ($R$) & $\mu$
\\\br
& & & & & \\
$\mathcal{A}$ &  $\left\{\begin{tabular}{l} $n \neq 2$ \\
$n=2$ \end{tabular} \right.$ &
 \begin{tabular}{l} $a_0 |t|^\frac{(1-n)(2n-1)}{(n-2)}$ \\ $a_0 e^{\frac{1}{3} \Th_0 t}$  \end{tabular}& \ $0$ &
 \begin{tabular}{l} $\frac{6n(1-n)(2n-1)(4n-5)}{(n-2)^2 t^{2}}$ \\ $\frac{4}{3} \Th_0^2$ \end{tabular} &
 \ $0$\\

$\mathcal{B}$ &  \ $n=\frac{5}{4},~w=\frac{2}{3}$ & \ $a_0
|t|^\frac{1}{2} $ & \ $0$ & \ $0$ & \ $0$
\\

$\mathcal{C}$ & $\left\{\begin{tabular}{l} $w \neq -1$ \\
$w=-1$ \end{tabular} \right.$ &  \begin{tabular}{l} $a_0
|t|^\frac{2n}{3(1+w)}$\\ $a_0 e^{\frac{1}{3} \Th_0 t}$
\end{tabular} & \ $0$ & \begin{tabular}{l} $
\frac{4n(4n-3(1+w))}{3(1+w)^2t^{2}}$ \\ $\frac{4}{3} \Th_0^2$
\end{tabular} & \begin{tabular}{l} $\mu^\mathcal{C}_0 \;t^{-2n}$ \\ $\mu^\mathcal{C}_1$\end{tabular} \\

$\mathcal{F}$ & $\left\{\begin{tabular}{l} $n \neq 2$ \\
$n=2$ \end{tabular} \right.$ & \begin{tabular}{l} $a_0
|t|^\frac{(2n^2-5n+5)}{3(2-n)} $ \\ $a_0 e^{\frac{1}{3} \Th_0 t}$
\end{tabular} &
\begin{tabular}{l} $\frac{(1+2n-2n^2)}{\sqrt{3}(n-2)t}$ \\ $\frac{1}{\sqrt{3}} \Th_0$ \end{tabular} &
\begin{tabular}{l}$\frac{6n(1-n)(4n^2-10n+7)}{(n-2)^2 t^{2}}$ \\ $\frac{10}{3} \Th_0^2$
\end{tabular} & \ $0$
\\

$\mathcal{G}$ & $\left\{\begin{tabular}{l} $w \neq -1$ \\
$w=-1$ \end{tabular} \right.$ &  \begin{tabular}{l} $a_0
|t|^\frac{2n}{3(1+w)}$\\ $a_0 e^{\frac{1}{3} \Th_0 t}$
\end{tabular} &
\begin{tabular}{l} $\frac{(3(1+w)-2n)}{\sqrt{3}(1+w)t}$ \\ $\frac{1}{\sqrt{3}} \Th_0$ \end{tabular}
& \begin{tabular}{l} $ \frac{4n(2(n-1) +w(1+3w-2n))}{(1+w)^2t^{2}}$
\\ $\frac{10}{3} \Th_0^2$
\end{tabular} & \begin{tabular}{l} $\mu^\mathcal{G}_0 \;t^{-2n}$ \\
$\mu^\mathcal{G}_1$\end{tabular}\\
Line & & & & & \\

 $\mathcal{L}_{1}$ & $\Sigma_*^2= \frac{5-4n}{2n-1}$,  $n\in(1,\frac{5}{4})$&
\ \ $a_0 |t|^\frac{1}{2+\Sigma_*^2}$ & $\frac{\sqrt{3}
|\Sigma_*|}{(2+\Sigma_*^2)t}  $ & \ \ $0$ & \ \ $0$\\

$\mathcal{L}_{2}$ & $Q_*^2= {\frac{2n-1}{5-4n}}$,
$n\in(\frac{1}{2},1)$ &
\ \ $a_0 |t|^\frac{Q_*^2}{1+2Q_*^2}$ & $\frac{ \sqrt{3} |Q_*|}{(1+2Q_*^2)t}$ &\ \  $0$ & \ \ $0$\\
& & & & & \\
\br
\end{tabular}\label{tab:sols}
\end{table}


\begin{table}[tbp] \centering
\caption{Deceleration parameter for the \equi points. In the last
three columns we state explicitly for which values of $n$ the
deceleration parameter $q$ (stated in the second column) is less,
equal to or larger than $0$, i.e. whether the have accelerated,
de--Sitter--like or decelerated behaviours. The parameters are
$P_+=\frac{1}{2}(1+\sqrt{3})$ and $S_w=\sfrac{3}{2}(1+w)$.}
\begin{tabular}{lccccc}
\multicolumn{6}{c}{}\\
\br  &   &  &  $q=-1$ & $-1<q<0$ & $q>0$
\\\cline{4-6}
Point  &  $q$ & $w$ & \multicolumn{3}{c}{Range of $n$}     \\\br
$\mathcal{A}$ &
 $\frac{1+2n-2n^2}{1-3n+2n^2}$ & all & $2$
  & $(P_+,2)$ & $\left\{\begin{tabular}{c} $\left(0,\sfrac{1}{2}\right)$ \\ $(1,P_+)$ \end{tabular}
 \right.$\\
$\mathcal{B}$ & $1$ & all & -  & - & $(0,\infty)$\\
$\mathcal{C}$ &  $ \frac{3(1+w)-2n}{2n}$ & \begin{tabular}{c} $-1$\\
$[0,1]$ \end{tabular}&  \begin{tabular}{c}$(0,\infty)$\\ -
\end{tabular} & \begin{tabular}{c}- \\ $(S_w,\infty)$
\end{tabular} & \begin{tabular}{c} - \\ $(0,S_w)$
\end{tabular} \\
 $\mathcal{F}$ &   $\frac{1+2n-2n^2}{5-5n+2n^2} $ & all & - &
$(P_+,2)$ & $(0,P_+)$ \\
$\mathcal{G}$ &  $ \frac{3(1+w)-2n}{2n}$ & \begin{tabular}{c} $-1$\\
$[0,1]$ \end{tabular} & \begin{tabular}{c}$(0,\infty)$\\ -
\end{tabular} & \begin{tabular}{c}- \\
$(S_w,\infty)$
\end{tabular} & \begin{tabular}{c} - \\ $(0,S_w)$
\end{tabular} \\
Line & & & & &\\
 $\mathcal{L}_{1}$ &
$1+\Sigma_*^2$ & all &- & $(0,\infty)$ & -\\
$\mathcal{L}_{2}$ &
$1+\frac{1}{Q_*^2}$ & all &- & $(0,\infty)$ & -\\
\br
\end{tabular}\label{tab:q}
\end{table}

\subsection{The full state space}\label{Subsec:state-space}
The full state space is obtained by matching the various sectors
along their common boundaries. Because the full state space is
4--dimensional it is  not easily visualised, so we refer to
\cite{Goheer07b} for an illustration of the 2--dimensional Bianchi I
vacuum subspace and the  2--dimensional flat FLRW subspace with
matter. We emphasise that we have to formally exclude the subset
with $Q=0$ and $x\neq 0$ from the state space unless $y\rightarrow
0$. This is an artifact of the definition of the variable $x$, and
reflected by the fact that there are no orbits crossing this subset
-- the only trajectories crossing the plane $Q=0$ pass through the
points with $x=0$ or $y=0$.

\subsection{Qualitative Analysis}\label{Subsec:Qual}
We summarise the dynamical behaviour of the equilibrium points and
lines of \equi \  points in Tables \ref{tab:nature-B1},
\ref{tab:nature-B3} and \ref{tab:nature-KS} and Tables
\ref{tab:nature-L1} and \ref{tab:nature-L2} respectively. For the
stability analysis, we only consider the four cases: cosmological
constant $w=-1$, dust $w=0$, radiation $w=1/3$ and stiff matter
$w=1$. We only state the results for the \equi
 points corresponding to expanding solutions. The collapsing points
 are obtained by time--reversal -- in other words their dynamical
 stability properties are simply reversed: If $\mathcal{A}_+$ is a
 repeller for a given range of $n$, then $\mathcal{A}_-$ is an
 attractor for the same range of $n$.

 Table \ref{tab:nature-B1} consists of all the BI subspace equilibrium
 points (excluding the lines); their behaviour is similar to the flat Friedmann points
 which were found previously \cite{Leach06,Carloni05}.  We note that
 some of the solutions corresponding to the Friedmann and BI \equi points, have been found in \cite{Barrow06,Clifton06a}.

 The lines of equilibrium points have to be treated more carefully.
 We summarise their dynamical behaviour in Tables \ref{tab:nature-L1}
 and \ref{tab:nature-L2}. As noted above, the two lines include the
 same parametric solutions, but $\mathcal{L}_{1}$ corresponds to $x\geq 0$,
 while $\mathcal{L}_{2}$ has $x\leq 0$.

Since these lines correspond to flat solutions they should have been
found in \cite{Leach06}. In fact, the authors of \cite{Leach06}
found a line of \equi   points denoted by $\mathcal{L}^*_{1}$,
extending over $\Sigma_{**}\in[0,\,\infty)$, where $\Sigma_{**}$
measures the shear contribution to the Friedmann equation. The range
$0\leq\Sigma_{**}\leq 1$ corresponds to our $\mathcal{L}_{1,+}$,
while $\Sigma_{**}>1$ corresponds to $\mathcal{L}_{2,+}$ as can
be seen from the solutions of the scale factor in \cite{Leach06}. In
\cite{Leach06} the isotropic solution was at
 $\Sigma_{**}=0$ and the static one occured for $\Sigma_{**}\rightarrow \infty$.
 Note that \cite{Leach06} did not address the collapsing solutions
$\mathcal{L}_{1,-}$ or opposite orientation of the shear
$\mathcal{L}_{2,-}$ because the phase space for BI is symmetric
about the plane $\Sigma=0$. The stability for $\Sigma<0$ can be
obtained by time reversal from the corresponding points in the
$\Sigma>0$ subspace.

As stated above, $\mathcal{L}_{1,2}$ are only physical for certain
special coordinates. These are
\begin{equation*}
\Sigma_*=\pm\sqrt{\frac{5-4n}{2n-1}}, \ \  n\in (1,5/4), \ \ {\rm
and} \ \  Q_*=\pm\sqrt{\frac{2n-1}{5-4n}}, \ \ n\in (1/2,1),
\end{equation*}
for $\mathcal{L}_{1}$ and $\mathcal{L}_{2}$ respectively. These
\equi points are always saddles in nature.

The nature of the BIII \equi points is stated in Table
\ref{tab:nature-B3}. For the sake of completeness we have included
the stability of points $\mathcal{D}$ and $\mathcal{E}$, but will
not discuss them any further since they are not physical. The point
$\mathcal{F}$ lies in the BIII subspace for
$n\in(0,\frac{1}{2}(1+\sqrt{3}))$. $\mathcal{F}_+$ is a saddle for
$w=-1$ and for $n\in (1,5/4)$ when $w=0$, but an attractor
otherwise. $\mathcal{G}$ lies in the BIII subspace for $n\in(0,3/2)$
when $w=0$, for all $n$ when $w=1/3$ and $n\in(0,1)$ and
$n\in(3,\infty)$ when $w=1$. $\mathcal{G}_+$ is saddle except for
$n\in (1,5/4)$ when $w=0$ where it becomes an attractor.

The nature of the KS \equi points is stated in Table
\ref{tab:nature-KS}. Point $\mathcal{F}$ lies in the KS subspace for
$n>\frac{1}{2}(1+\sqrt{3})$ and $\mathcal{F}_+$ is always a saddle.
Similarly, $\mathcal{G}$ lies in the KS subspace for all $n$ when
$w=-1$, for $n>3/2$ when $w=0$ and $n\in(1,3)$ when $w=1$.
$\mathcal{G}_+$ is saddle except for $n\in (1,1.13)$ when $w=1$,
where it is a repeller.

We can identify the following global attractors and repellers:
$\mathcal{A}_+$ is a global attractor for $n\in (P_+,2)$ when $w=0$,
$1/3$ and $1$, and for $n\in (2,\infty)$ (all $w$). When $w=-1$,
$\mathcal{C}_+$ is a global attractor for $n\in (1,2)$ and
$\mathcal{E}_+$ for $n\in (0,1/2)$. Point $\mathcal{F}_+$ is a
global attractor for $n\in (0,1/2)$ and $n\in (5/4,P_+)$ when $w=0$,
$1/3$ and $1$, and for $n\in (1,5/4)$ when $w=1/3$ and $1$.
$\mathcal{G}_+$ is only a global attractor for  $n\in (1,5/4)$ when
$w=0$. By time reversal the corresponding contracting solutions are
global repellers. There are no global repellers in the expanding
subspace since the lines $\mathcal{L}_{1,2}$ contain repellers and
hence there are no global attractors in the collapsing subspace.

\begin{table}[tbp] \centering
\caption{Nature of the expanding ($\eps=+1$) spatially flat BI \equi
points. The collapsing analogs are simply time reversed. The
parameters are $P_+=\frac{1}{2}(1+\sqrt{3})$ and
$V_+=\frac{1}{14}(11+\sqrt{37})$.}
\begin{tabular}{cccccccccc}
\multicolumn{10}{c}{} \\ \br
 \footnotesize{Point} & \footnotesize{$w$} & \multicolumn{8}{c}{\footnotesize{Range of n}}\\
 \cline{3-10}
 & &  \footnotesize{$(0,1/2)$} & \footnotesize{$(1/2,1)$} & \footnotesize{$(1, V_+)$}
& \footnotesize{$(V_+,5/4)$} & \footnotesize{$(5/4,P_+)$} &
\footnotesize{$(P_+,3/2)$} &
\footnotesize{$(3/2,2)$} & \footnotesize{$(2,\infty)$}\\
\hline

\hline \hline

\footnotesize{$\mathcal{A}_+$}& \footnotesize{$-1$} &
\footnotesize{Saddle} & \footnotesize{Attractor}
&\footnotesize{Repeller} & \footnotesize{Repeller} &
\footnotesize{Saddle} & \footnotesize{Saddle} & \footnotesize{Saddle} & \footnotesize{Attractor} \\

&\footnotesize{$0$} & \footnotesize{Saddle}
&\footnotesize{Attractor} & \footnotesize{Repeller} &
\footnotesize{Repeller} & \footnotesize{Saddle} &
\footnotesize{Attractor} & \footnotesize{Attractor} &
\footnotesize{Attractor}\\

& \footnotesize{$1/3$} & \footnotesize{Saddle} &
\footnotesize{Attractor} & \footnotesize{Repeller} &
\footnotesize{Repeller} &
\footnotesize{Saddle} &  \footnotesize{Attractor} &  \footnotesize{Attractor} &  \footnotesize{Attractor} \\

& \footnotesize{$1$} & \footnotesize{Saddle} &
\footnotesize{Attractor} & \footnotesize{Repeller} &
\footnotesize{Saddle} & \footnotesize{Saddle} & \footnotesize{Attractor} & \footnotesize{Attractor} & \footnotesize{Attractor}\\
\hline

\small{$\mathcal{B}_+$} & \footnotesize{$-1$} &
\footnotesize{Saddle} & \footnotesize{Saddle} &
\footnotesize{Saddle} & \footnotesize{Saddle} &
\footnotesize{Saddle} & \footnotesize{Saddle} & \footnotesize{Saddle} & \footnotesize{Saddle}\\

& \footnotesize{$0$} & \footnotesize{Saddle} & \footnotesize{Saddle}
& \footnotesize{Saddle} & \footnotesize{Saddle} &
\footnotesize{Saddle} & \footnotesize{Saddle} & \footnotesize{Saddle} & \footnotesize{Saddle}\\

& \footnotesize{$1/3$} & \footnotesize{Saddle} &
\footnotesize{Saddle} & \footnotesize{Saddle} &
\footnotesize{Saddle} &
\footnotesize{Saddle} & \footnotesize{Saddle} & \footnotesize{Saddle} & \footnotesize{Saddle}\\

 & \footnotesize{$1$} &
\footnotesize{Repeller} & \footnotesize{Repeller} &
\footnotesize{Saddle} & \footnotesize{Saddle} &
\footnotesize{Saddle} & \footnotesize{Saddle} &
\footnotesize{Repeller} &
\footnotesize{Repeller}\\
\hline

\small{$\mathcal{C}_+$} & \footnotesize{$-1$} &
\footnotesize{Saddle} & \footnotesize{Saddle} &
\footnotesize{Attractor} & \footnotesize{Attractor} &
\footnotesize{Attractor}
& \footnotesize{Attractor} & \footnotesize{Attractor} &\footnotesize{Saddle} \\

& \footnotesize{$0$} & \footnotesize{Saddle} & \footnotesize{Saddle}
& \footnotesize{Saddle} & \footnotesize{Saddle} &
\footnotesize{Saddle} & \footnotesize{Saddle} & \footnotesize{Saddle} & \footnotesize{Saddle}\\

& \footnotesize{$1/3$} & \footnotesize{Saddle} &
\footnotesize{Saddle} & \footnotesize{Saddle} &
\footnotesize{Saddle} &
\footnotesize{Saddle} & \footnotesize{Saddle} & \footnotesize{Saddle} & \footnotesize{Saddle}\\

& \footnotesize{$1$} & \footnotesize{Saddle} & \footnotesize{Saddle}
& \footnotesize{Saddle} & \footnotesize{Repeller} &
\footnotesize{Repeller} & \footnotesize{Repeller} &
\footnotesize{Saddle} &
\footnotesize{Saddle}\\
\br
\end{tabular}\label{tab:nature-B1}
\end{table}

\begin{table}[tbp] \centering
\caption{Nature of the line of expanding spatially flat anisotropic
equilibrium points $\mathcal{L}_{1+}$. Here
$\Sigma_b(n)=\sqrt{\sfrac{5-4n}{2n-1}}$ is a bifurcation value
depending on $n$.}
\begin{tabular}{cccccc}
\multicolumn{5}{c}{} & \\ \br
 \footnotesize{$w$} & \footnotesize{$n$} &\footnotesize{$\Sigma \in[-1,-\Sigma_b(n))$} &
 \footnotesize{$\Sigma \in(-\Sigma_b(n),\Sigma_b(n))$} & \footnotesize{$\Sigma \in(\Sigma_b(n),1)$} \\
\hline

&\footnotesize{$n\in(0,1)$} & \footnotesize{Repeller} & \footnotesize{Repeller} & \footnotesize{Repeller}\\

\footnotesize{$-1,0,1/3$}&\footnotesize{$n\in(1,5/4)$}&
\footnotesize{Repeller}
&\footnotesize{Saddle}& \footnotesize{Repeller}\\

&\footnotesize{$n>5/4$}& \footnotesize{Repeller} & \footnotesize{Repeller} & \footnotesize{Repeller}\\
\hline
\footnotesize{$1$} & \footnotesize{All $n$} & \footnotesize{Saddle} & \footnotesize{Saddle} & \footnotesize{Saddle}\\
\br
\end{tabular}\label{tab:nature-L1}
\end{table}

\begin{table}[tbp] \centering
\caption{Nature of the line of spatially flat anisotropic
equilibrium points $\mathcal{L}_{2,\pm}$. Here
$Q_b(n)=\sqrt{\sfrac{2n-1}{5-4n}}$ is a bifurcation value depending
on $n$. We discuss the  bifurcation $Q=0$ in the section on static
solutions below. Note that the dynamical behaviour of
$\mathcal{L}_{2,+}$ and $\mathcal{L}_{2,-}$ is identical.}
\begin{tabular}{cccccc}
\multicolumn{5}{c}{} & \\ \br \footnotesize{$n$}  & \footnotesize{
$Q \in[-1,-Q_b(n))$} & \footnotesize{$Q \in(-Q_b(n),0)$} &
 \footnotesize{$Q \in(0,Q_b(n))$} & \footnotesize{$Q \in(Q_b(n),1]$} \\
\hline

\footnotesize{$n\in[0,1/2]$} & \footnotesize{Attractor}  &
\footnotesize{Attractor} & \footnotesize{Repeller}
 & \footnotesize{Repeller}\\

\footnotesize{$n\in(1/2,1)$}& \footnotesize{Attractor}
& \footnotesize{Saddle} & \footnotesize{Saddle} & \footnotesize{Repeller}\\

\footnotesize{$n>1$}& \footnotesize{Attractor}  &
\footnotesize{Attractor} & \footnotesize{Repeller}
 & \footnotesize{Repeller}\\
\br
\end{tabular}\label{tab:nature-L2}
\end{table}

\begin{table}[tbp] \centering
 \caption{Nature of the spatially open Bianchi III  \equi points, where $P_+=\frac{1}{2}(1+\sqrt{3})$.}
\vspace{5mm}
\begin{tabular}{cccccccccc}
\multicolumn{7}{c}{} & \\ \br
\footnotesize{Point} & \footnotesize{$w$} & \multicolumn{6}{c}{\footnotesize{range of n}}\\
\cline{3-8} & &  \footnotesize{$(0,1)$} & \footnotesize{$(1,5/4)$} &
\footnotesize{$(5/4, P_+)$} & \footnotesize{$(P_+,3/2)$} &
\footnotesize{$(3/2,3)$} & \footnotesize{$(3,\infty)$}\\
\hline
\small{$\mathcal{D}_+$} & \footnotesize{All} & \footnotesize{Saddle}
& \footnotesize{Saddle} & \footnotesize{Saddle} &
\footnotesize{Saddle}
& \footnotesize{Saddle}  & \footnotesize{Saddle}\\
\hline

\small{$\mathcal{E}_+$} & \footnotesize{$-1$}
&\footnotesize{Attractor}& \footnotesize{Saddle} &
\footnotesize{Saddle} & \footnotesize{Saddle}
& \footnotesize{Saddle}  & \footnotesize{Saddle}\\
&\footnotesize{$0,\,1/3,\,1$} &\footnotesize{Saddle} &
\footnotesize{Saddle} & \footnotesize{Saddle} &
\footnotesize{Saddle}
& \footnotesize{Saddle}  & \footnotesize{Saddle}\\

\hline \footnotesize{$\mathcal{F}_+$}& \footnotesize{$-1$} &
\footnotesize{Saddle} & \footnotesize{Saddle} &
\footnotesize{Saddle} & - & -& - \\
&\footnotesize{$0$} & \footnotesize{Attractor} &
\footnotesize{Saddle} &
\footnotesize{Attractor} & - & -& - \\
& \footnotesize{$1/3$} & \footnotesize{Attractor}  & \footnotesize{Attractor} & \footnotesize{Attractor} & - & -& - \\
& \footnotesize{$1$} & \footnotesize{Attractor}  & \footnotesize{Attractor} & \footnotesize{Attractor} & - & -& - \\

\hline \small{$\mathcal{G}_+$}
& \footnotesize{$-1$} & - & -& - & - & -& -  \\
&\footnotesize{$0$} & \footnotesize{Saddle}
&\footnotesize{Attractor} & \footnotesize{Saddle}
&\footnotesize{Saddle} &  -& - \\

 & \footnotesize{$1/3$} &
\footnotesize{Saddle} & \footnotesize{Saddle} &
\footnotesize{Saddle} & \footnotesize{Saddle}
& \footnotesize{Saddle}  & \footnotesize{Saddle}\\
& \footnotesize{$1$} & \footnotesize{Saddle}&-&-&-&-&\footnotesize{Saddle}\\
\br
\end{tabular}\label{tab:nature-B3}
\end{table}

\begin{table}[t] \centering
 \caption{Nature of the spatially closed Kantowski-Sachs \equi
points, where $P_+=\frac{1}{2}(1+\sqrt{3})$ and $X\approx 1.13$.}
\vspace{5mm}
\begin{tabular}{cccccccc}
\multicolumn{8}{c}{} \\ \br
\footnotesize{Point} & \footnotesize{$w$} & \multicolumn{6}{c}{\footnotesize{Range of n}}\\
 \cline{3-8}
 & &  \footnotesize{$(0,1)$} & \footnotesize{$(1,X)$} & \footnotesize{$(X, P_+)$}
& \footnotesize{$(P_+,3/2)$} &
\footnotesize{$(3/2,3)$} & \footnotesize{$(3,\infty)$}\\
\hline
\footnotesize{$\mathcal{F}_+$}& \footnotesize{$-1$} & - & -& - &
\footnotesize{Saddle} & \footnotesize{Saddle}
& \footnotesize{Saddle} \\
&\footnotesize{$0$} & - & -& - & \footnotesize{Saddle} &
\footnotesize{Saddle}
& \footnotesize{Saddle} \\
& \footnotesize{$1/3$} & - & -& - & \footnotesize{Saddle} &
\footnotesize{Saddle}
& \footnotesize{Saddle} \\
& \footnotesize{$1$} & - & -& - & \footnotesize{Saddle} &
\footnotesize{Saddle}
& \footnotesize{Saddle} \\
\hline

\small{$\mathcal{G}_+$} & \footnotesize{$-1$} &
\footnotesize{Saddle} & \footnotesize{Saddle} &
\footnotesize{Saddle} & \footnotesize{Saddle} &
\footnotesize{Saddle}
& \footnotesize{Saddle} \\
&\footnotesize{$0$}  & - & -& - & -& \footnotesize{Saddle} & \footnotesize{Saddle}  \\
& \footnotesize{$1/3$} & - & -& - & -& -& - \\
& \footnotesize{$1$} & - & \footnotesize{Repeller} & \footnotesize{Saddle} & \footnotesize{Saddle} & \footnotesize{Saddle} & - \\
\br
\end{tabular}\label{tab:nature-KS}
\end{table}

\subsection{Bouncing or recollapsing trajectories}\label{Subsec:bounces}
As motivated above, any trajectory corresponding to a bouncing or
recollapsing solution must pass through $x=Q=0$ or $y=Q=0$.

The existence of bouncing orbits for Bianchi I models has been
studied in \cite{Goheer07b}. In the vacuum case it was found that
there exist bouncing/recollapsing trajectories, but only for $y<0$.
If $n>1$, $R$ has to be negative and there can only be re-collapse
($\dot{\Theta}<0$). For $n\in [0,1/2]$ re-collapse may occur if
$R>0$, and for $n\in [0,1/2]$ there may be a bounce
($\dot{\Theta}>0$) for positive $R$. In all cases, the  bouncing
trajectories have to pass through the single point  $x=Q=0$ (denoted
by $\tilde{\mathcal M}$  in \cite{Goheer07b}) in the 2-dimensional BI
 vacuum subspace. Note that it is not possible to achieve a bounce
 through $y=Q=0$ here, since a line of \equi points
 passes through that point  in this subspace.

When matter is added, we obtain another degree of freedom, and
unlike in GR, the matter term may enhance bouncing or recollapsing
behaviour due to the $R^{n-1}$ term coupled to the energy density.
The corresponding trajectories now have to pass through the
1-dimensional lines with $x=Q=0$ or $y=Q=0$ instead of the single
point $\tilde{\mathcal M}$.

In the presence of spatial curvature, it is yet easier to achieve
bouncing or recollapsing behaviour.  If ${}^3R<0$, the results from
the flat Bianchi I case are qualitatively recovered. For ${}^3R>0$
however, there are differences to the Bianchi I case. In particular,
positive spatial curvature allows $\Theta=0$ even for positive $y$.

\section{Isotropisation of OSH Bianchi models}

It is possible to study isotropisation by looking at the stability
of the Friedmann points in the state space (see \cite{Wainwright04}
and references therein). When such an isotropic point is an
attractor, then we have asymptotic isotropisation in the future. If
the point is a repeller we have an isotropic initial singularity,
and when it is a saddle we have intermediate isotropisation. Because
of the dimensionality and complexity of the state space, we will not
study specific orbits to investigate viable models. We will
therefore restrict the following discussion to the behaviour around
the equilibrium points only.

In the previous sections we found two isotropic \equi points that
admit cosmological solutions: a vacuum point $\mathcal{A}$ and a
non-vacuum point $\mathcal{C}$. These points were also found in the
BI case \cite{Leach06}.

The expanding point $\mathcal{A}_+$ is an isotropic \emph{past}
attractor for $n\in (1,5/4)$ when $w=-1$, $0$ or $1/3$, and for
$n\in (1,\sfrac{1}{14}(11+\sqrt{37}))$ when $w=1$. As pointed out in
\cite{Leach06}, this is an interesting feature, since the existence
of an isotropic past attractor implies that we do not require
special initial conditions for inflation to take place. The
contracting analog $\mathcal{A}_-$ is an isotropic future attractor
in these ranges. $\mathcal{A}_+$ is a \emph{future} attractor for
$n>2$ when $w=-1$ and for $n>\sfrac{1}{2}(1+\sqrt{3})$ when $w=0$,
$1/3$ or $1$. By time reversal, $\mathcal{A}_-$ is a past attractor
for these parameter values.

The \equi point $\mathcal{C}_+$ is an isotropic past attractor for
$n\in (\sfrac{1}{14}(11+\sqrt{37},3/2)$ when $w=1$ and an isotropic
future attractor for $n\in (1,2)$ when $w=-1$. When $w=0$ or $w=1/3$
this point is a saddle for all values of $n$. This means that in
this case we have a transient matter/ radiation dominated phase in
which the model is highly isotropic and hence potentially compatible
with observations.

We note that all isotropic \equi points found in this analysis are
flat Friedmann like, unlike in  \cite{Carloni05}, where the
isotropic points $\mathcal{A}$ and $\mathcal{C}$ with non-zero
spatial curvature  were found. The reason for this discrepancy is
that the plane $\Sigma=0$ is no longer invariant when allowing for
non--zero spatial curvature ($k\neq0$); as in GR spatial curvature
causes anisotropies to grow in models with $R^n$-gravity. For this
reason the points $\mathcal{A}$ and $\mathcal{C}$ no longer remain
\equi points in the full OSH Bianchi state space.

There are two \equi points of interest with non-zero shear: the
vacuum point $\mathcal{F}$ and the non-vacuum point $\mathcal{G}$.
These points are isotropic for certain bifurcation values of $n$ and
$w$: $\mathcal{F}$ is isotropic for $n=\sfrac{1}{2}(1+\sqrt{3})$ for
all $w$, and $\mathcal{G}$ is isotropic for $n=\sfrac{3}{2}(1+w)$ if
$w>-1$. The KS point $\mathcal{G}_+$ is a past attractor for
$n\in(1,1.13)$ when $w=1$ and a saddle for $n>3/2$ when $w=0$. This
means that we can have initial conditions which are anisotropic, or
we can have intermediate anisotropic conditions which are conducive
for structure formation, provided that the anisotropies are
sufficiently small. When $w=1/3$, the point $\mathcal{G}_+$ lies in
the BIII state space and is a saddle for all values of $n$, and when
$w=0$ the same applies for $n\in(0,1)$ or $n\in (5/4,3/2)$.

\section{Remarks and Conclusions}

Our main aim in this paper was to investigate the effects of spatial
curvature on the isotropisation of OSH Bianchi models in
$R^n$-gravity, and to possibly identify static solutions and bounce
behaviours. To achieve this goal, we constructed a compact state
space which allows one to obtain a complete picture of the
cosmological behaviour for expanding, contracting and static as well
as bouncing or recollapsing models. This is not possible with the
non-compact variables used in \cite{Leach06}, since the \equi points
with static solutions do not have finite coordinates in this
framework. The Poincar\'{e} projection also does not allow one to
patch together the expanding and contracting copies of the state
space, so bounce behaviour cannot be investigated. This is discussed
in detail in \cite{Goheer07b} for the BI subspace, where the results
obtained here are compared to the results obtained in
\cite{Leach06}.

We do not find any exact Einstein static solutions in this analysis.
However we do find orbits that exhibit cyclic behaviour, which was
expected from previous work examining the conditions for bouncing
solutions in $f(R)$ gravity \cite{Carloni06}. We also recover all
the isotropic \equi points that were found in \cite{Leach06}. The
expanding vacuum point $\mathcal{A}_+$ is a past attractor for $n\in
(1,5/4)$ as in the BI case. We emphasise that we only find
\emph{flat} ($k=0$) isotropic \equi points ($\mathcal{A}$,
$\mathcal{B}$ and $\mathcal{C}$).  Therefore for these types of
theories, isotropisation also implies cosmological behaviours which
evolve towards spatially flat spacetimes. Late time behaviour with
non-zero spatial curvature will  have a growth in anisotropies, as
in GR.

In conclusion, we have shown that spatial curvature does indeed
affect the isotropisation of cosmological models in $R^n$-gravity.
While no exact static solutions could be found, we did find that
bounces can occur in these cosmologies.

\ack

This research was supported by the National Research Foundation
(South Africa) and the Italian {\it Ministero Degli Affari Esteri-DG
per la Promozione e Cooperazione Culturale} under the joint Italy/
South Africa Science and Technology agreement. NG is funded by the
Claude Leon Foundation. We thank Salvatore Capozziello for useful
discussions.

\section*{References}


\begin{thebibliography}{99}

\bibitem{DEfR} Carroll S M, Duvvuri V, Trodden M and Turner M S
2004 \prd {\bf 70} 043528; Nojiri S and Odintsov S D 2003 \prd {\bf
68} 123512; Capozziello S 2002 {\it Int. Journ. Mod. Phys.} D {\bf
11} 483; Salti M arXiv: gr-qc/0607116"; Faraoni V 2005 \prd {\bf 72}
124005; Ruggiero M L and Iorio L 2007 JCAP {\bf 0701} 010; de la
Cruz-Dombriz A and Dobado A 2006 \prd {\bf 74} 087501; Poplawski N J
2006 \prd {\bf 74} 084032; Poplawski N J 2007 \cqg {\bf 24} 3013;
Brookfield A W, van de Bruck C and Hall L M H 2006 \prd {\bf 74}
064028; Song, Y, Hu W and Sawicki I 2007 \prd {\bf 75} 044004; Li B,
Chan K and Chu M 2007 \prd {\bf 76} 024002; Jin X, Liu D and Li X
arXiv: astro-ph/0610854; Sotiriou T P and Liberati S 2007  \aph {\bf
322} 935; Sotiriou T P 2006 \cqg {\bf 23} 5117; Bean R, Bernat D,
Pogosian L, Silvestri A and Trodden M 2007 \prd {\bf 75} 064020;
Navarro I and Van Acoleyen K 2007 JCAP {\bf 0702} 022; Bustelo A J
and Barraco D E 2007 \cqg {\bf 24} 2333 Olmo G J 2007 \prd {\bf 75}
023511; Ford J, Giusto S and Saxena A arXiv: hep-th/0612227;
Briscese F, Elizalde E, Nojiri S and Odintsov S D 2007 \plb {\bf
646} 105; Baghram S, Farhang M and Rahvar S  2007 \prd {\bf 75}
044024; Bazeia D, Carneiro da Cunha B, Menezes R and Petrov A 2007
\plb {\bf 649} 445; Zhang P 2007 \prd {\bf 76} 024007; Li B and
Barrow J D 2007 \prd {bf 75} 084010; Rador T 2007 \plb {\bf 652}
228; Rador T 2007 \prd {\bf 75} 064033; Sokolowski L M arXiv:
gr-qc/0702097; Faraoni V 2007 \prd {\bf 75} 067302; Bertolami O,
Boehmer C G, Harko T and Lobo F S N 2007 \prd {\bf 75} 104016;
Srivastava S K arXiv:0706.0410 [hep-th]; Capozziello S, Cardone V F
and Troisi A 2006 JCAP {\bf 08} 001; Starobinsky A A arXiv:
0706.2041 [gr-qc]

\bibitem{Nojiri06}
Nojiri S and Odintsov S D 2006  \prd {\bf 74} 086005

\bibitem{Capozzi07b} Capozziello S, Cardone V F and Troisi A 2007 \mn {\bf 375} 1423

\bibitem{Nojiri07} Nojiri S and Odintsov S D 2007 \plb {\bf 652} 343

\bibitem{Leach06} Leach J A, Carloni S and Dunsby P K S 2006 \cqg {\bf 23} 4915

\bibitem{Barrow06c} Barrow J D and Hervik S 2006 \prd {\bf 74}
124017

\bibitem{Barrow07a} Barrow J D and Middleton J \prd {\bf 75} 123515

\bibitem{Wald83} Wald R M 1983 \prd {\bf 28} 2118

\bibitem{Goode85} Goode S W and Wainwright J 1985 \cqg {\bf 2} 99

\bibitem{vElst96}  van Elst H and Ellis G F R 1996 \cqg {\bf
13} 1099

\bibitem{Ellis67} Ellis G F R 1967 \jmp {\bf 8} 1171

\bibitem{Stewart68} Stewart J M and Ellis G F R 1968 \jmp {\bf 9}
1072

\bibitem{Goliath99} Goliath M and Ellis G F R 1999 \prd {\bf 60}
023502

\bibitem{DSCosmo} Coley A A 2003 {\it
Dynamical systems and cosmology} (Dordrecht: Kluwer Academic
Publishers)

\bibitem{Clifton05b} Clifton T and Barrow J D 2005 \prd {\bf 72}
123003

\bibitem{Bohmer07} B\"{o}hmer C G, Hollenstein L and Lobo S N L arXiv:
0706.1663 [gr-qc]

\bibitem{Tolman}
Tolman R C 1934 {\it Relativity Thermodynamics and Cosmology}
(Oxford University Press, 1934; Dover, 1987).

\bibitem{Barrow03} Barrow J D, Ellis G F R, Maartens R and Tsagas C G 2003 \cqg {\bf 20}
L155

\bibitem{Carloni06} Carloni S, Dunsby P K S, and Solomons D 2006 \cqg {\bf 23}
1913

\bibitem{Clifton05} Clifton T and Barrow J D 2005 \prd {\bf 72} 103005

\bibitem{Clifton07} Clifton T 2007 \cqg {bf 24} 5073

\bibitem{Solomons06} Solomons D, Dunsby P K S and Ellis G F R  2006 \cqg {\bf 23}
6585

\bibitem{Dynamical} Wainwright J and Ellis G F R (ed) 1997 {\it
Dynamical systems in cosmology} (Cambridge: Cambridge University
Press) (see also references therein)

\bibitem{Wainwright04} Wainwright J and Lim W C 2005 J. Hyperbol.
Diff. Equat. {\bf 2} 437

\bibitem{Carloni05} Carloni S, Dunsby P K S, Capozziello S and Troisi
A 2005 \cqg {\bf 22} 4839

\bibitem{Carloni07} Carloni S, Leach J A, Capozziello S and  Dunsby P K S archiv:
gr-qc/0701009

\bibitem{Abdel07} Abdelwahab M, Carloni S, Dunsby P K S arXiv: 0706.1375 [gr-qc]

\bibitem{Carloni07a} Carloni S, Troisi A, Dunsby P K S arXiv:
0706.0452 [gr-qc]

\bibitem{Amendola07a} Amendola L, Gannouji R, Polarski D and
Tsujikawa S 2007 \prd {\bf 75} 083504

\bibitem{Amendola07b} Amendola L, Polarski D and
Tsujikawa S 2007 \prl {\bf 98} 131302

\bibitem{Cognola07} Cognola G, Gastaldi M, and
Zerbini S arXiv: gr-qc/0701138

\bibitem{Agarwal07} Agarwal N and Bean R arXiv:
0708.3967 [astro-ph]

\bibitem{Goheer07b} Goheer N, Leach J A and Dunsby P K S arXiv: 0710.0819 [gr-qc]

\bibitem{Poincare} Poincar\'{e} H 1881 {\it J. Math\'{e}matiques}
{\bf 7} 375

\bibitem{Perko} Perko L 1996 {\it Differential equations and
dynamical systems} (New York: Springer Verlag)

\bibitem{Campos01a} Campos A and Sopuerta 2001 \prd {\bf 63} 104012

\bibitem{Campos01b} Campos A and Sopuerta 2001 \prd {\bf 64} 104011

\bibitem{Dunsby04} Dunsby P K S, Goheer N, Bruni M and Coley A 2004 \prd {\bf
69} 101303

\bibitem{Goheer04} Goheer N, Dunsby P K S, Bruni M and Coley A 2004 \prd {\bf
70} 123517

\bibitem{Rippl96} Rippl S, van Elst H, Tavakol R and Taylor D 1996 \grg
\textbf{28} 193


\bibitem{Barrow06} Barrow J D and Clifton T 2006 \cqg {\bf
23} L1

\bibitem{Clifton06a} Clifton T and Barrow J D 2006 \cqg {\bf
23} 2951

\end{thebibliography}
\end{document}